 \documentclass[final,5p,times,twocolumn]{elsarticle}

\usepackage{hyperref}
\usepackage{comment}
\usepackage{ifpdf}
\usepackage{subfigure}
\usepackage{amssymb}
\usepackage{amsfonts}
\usepackage{epsf}
\usepackage{rotating}
\usepackage{graphicx}
\usepackage{amsmath}
\usepackage{fancyhdr}
\usepackage{lineno}
\usepackage{dutchcal}
\usepackage{graphics}
\usepackage{pstricks}
\usepackage{color}
\usepackage{multirow}
\usepackage[utf8]{inputenc} 
\usepackage{lineno}

\newcommand{\nn}{\nonumber}
\newcommand{\lsim}{\mathrel{\mathop{\kern 0pt \rlap
  {\raise.2ex\hbox{$<$}}}
  \lower.9ex\hbox{\kern-.190em $\sim$}}}
\newcommand{\gsim}{\mathrel{\mathop{\kern 0pt \rlap
  {\raise.2ex\hbox{$>$}}}
  \lower.9ex\hbox{\kern-.190em $\sim$}}}

\newcommand{\be}{\begin{equation}}
\newcommand{\ee}{\end{equation}}
\newcommand{\bea}{\begin{eqnarray}}
\newcommand{\eea}{\end{eqnarray}}

\biboptions{numbers,sort&compress}

\begin{document}

\begin{frontmatter}

\title{Theoretical constraints on the Higgs potential of the general $331$ model}

\author[label1]{Antonio Costantini}
\ead{antonio.costantini@bo.infn.it}

\author[label2]{Margherita Ghezzi}

\author[label3]{Giovanni Marco Pruna}

\address[label1]{INFN - Sezione di Bologna, Via Irnerio 46, 40126 Bologna, Italy}
\address[label2]{Institut f\"ur Theoretische Physik, Eberhard Karls Universit\"at T\"ubingen, Auf der Morgenstelle 14, 72076 T\"ubingen, Germany}
\address[label3]{INFN - Sezione di Roma Tre, Via della Vasca Navale 84, 00146 Rome, Italy}

\begin{abstract}
  This article reviews the theoretical constraints on the scalar potential of a general extension of the Standard Model that encompasses a $SU(3)_c\times SU(3)_L\times U(1)_X$ gauge symmetry. In this respect, the boundedness-from-below is analysed to identify the correct criteria for obtaining the physical minima of the Higgs parameter space. Furthermore, perturbativity and unitarity bounds are discussed in light of the exact diagonalisation of the scalar fields. 
  This study represents a framework for fast numerical checks on specific $331$ Model benchmarks that are relevant for future collider searches.
\end{abstract}


\end{frontmatter}

\section{Introduction}\label{sec:intro}
\noindent
The $331$ Model~\cite{Singer:1980sw,Valle:1983dk,Pisano:1991ee,Frampton:1992wt,Foot:1994ym,Hoang:1995vq} is an extension of the Standard Model (SM) where the non-abelian gauge group $SU(2)$ of the electroweak symmetry is promoted to an $SU(3)$, thus replacing the symmetry pattern from $SU(3)_c\times SU(2)_L \times U(1)_Y$ to $SU(3)_c\times SU(3)_L \times U(1)_X$. This assumption redefines the SM hypercharge as $\mathbb{Y}=\beta_Q \mathbb{T}^8+X\mathbb{I}$, where $\mathbb{T}^8$ is one of the well-known Gell--Mann matrix acting on $SU(3)_L$, $X$ is a new abelian charge assignment, $\mathbb{I}$ is the identity matrix and $\beta_Q$ is a free parameter of the model. When the latter is not specified, the setup is called ``general $331$ Model''.

Whilst each generation of the SM is anomaly free, in the general $331$ Model the cancellation of gauge anomalies occurs if and only if the total number of generations is (a multiple of) three. So far, all the experimental evidences seem to indicate that three is the number of active flavours in Nature. In contrast to the SM, where this fact represents a puzzling assumption, the general $331$ Model would represent an elegant and appealing solution to the mystery of flavour.

Concerning its particle spectrum, any realisation of the $331$ Model is accompanied by a rich variety of beyond-the-SM (BSM) particles that allows for heavy and potentially exotic states. Given the paucity of ``standard candles'' for new physics (NP) at the Large Hadron Collider (LHC) and low-energy experiments, a considerable attention was recently devoted to analysing exotic signatures of the $331$ Model in atomic physics~\cite{Long:2018fud}, flavour experiments~\cite{Buras:2012dp,Buras:2013dea,Buras:2016dxz}, and high-energy searches~\cite{Martinez:2014ltaGIUSTO,Hue:2015mna,Coriano:2018coq,RamirezBarreto:2019bpx}.

Even though different $331$ Model realisations provide very distinct phenomenological implications, all of them are characterised by the scalar potential detailed in Section~\ref{sec:model} of this paper. Remarkably, theoretical constraints on the scalar potential of such class of models were not systematically covered in previous literature. This article aims to fill this gap.

When studying the scalar potential of a model, one should derive a set of non-trivial conditions for its boundedness-from-below (BFB). Consequently, only the portion of the parameter space that fulfils them can be associated to existing vacua. In Section~\ref{sec:bfb}, this paper presents a set of necessary and sufficient conditions that applies to the case of the general $331$ Model, and can be easily implemented in future phenomenological studies.

Perturbative unitarity must be classified among the most important theoretical constraints to guarantee that perturbative scattering amplitudes at every order are related to a well-defined unitary $S$-matrix. This is especially important in BSM extensions with weak (rather than strong) new dynamics, as occurs in the general $331$ Model. In Section~\ref{sec:unitarity}, this paper describes how to organise the computation of the unitarity constraints in this framework.

Additional criteria should be fulfilled by physical parameters (masses and mixing angles) to ensure that every coupling of the scalar potential is perturbative, \emph{i.e.} less than or equal to $4\pi$. In Section~\ref{sec:perturbativity}, this paper shows that perturbativity constraints can place further conditions on the boson mass spectrum.

All the results presented in this article were obtained with the support of automated computational tools: both a SARAH~\cite{Staub:2013tta} and a FeynRules~\cite{Alloul:2013bka} model files were created and cross-checked against the model version already uploaded in the FeynRules Model Database~\cite{Cao:2016uur}; the FeynArts interface of FeynRules~\cite{Christensen:2009jx} was exploited to produce a model file for the FeynArts~\cite{Hahn:2000kx} and FormCalc~\cite{Hahn:1998yk,Hahn:2016ebn} packages; furthermore, the combined packages FeynArts and FormCalc were largely employed to study the perturbative unitarity constraints. All these results are collected in a dedicated Mathematica~\cite{Mathematica} file\footnote{Available on request.} to open the way for a fast selection of theoretically allowed portions of the general $331$ Model parameter space.

\section{The scalar sector of the general $331$ Model}\label{sec:model}
\noindent
The general $331$ Model represents a class of SM extensions containing the enlarged gauge group $SU(3)_c\times SU(3)_L\times U(1)_X$. Several specific versions can be obtained by a particular choice of the $\beta_{Q}$ parameter present in the definition of the electromagnetic charge operator:
\be
\mathbb{Q}=\mathbb{T}^3 + \beta_{Q}\mathbb{T}^8 + X\mathbb{I}.
\ee
In the following sections, the scalar sector of the $331$ Model will be explored without any specific assumption on $\beta_{Q}$. Besides, any details about the fermionic and gauge content of the model will be disregarded as can be found in dedicated literature~\cite{Buras:2012dp,Buras:2013dea,Hue:2017lak,Hung:2019jue}.

In the general $331$ Model, the electroweak symmetry breaking is realised by scalars accommodated within three triplets of $SU(3)_L$ 
\begin{equation}\label{eq:scalars}
\chi=\left(
\begin{array}{c}
\chi^A\\
\chi^B\\
\chi^0
\end{array}
\right)
,\quad\rho=\left(
\begin{array}{c}
\rho^+\\
\rho^0\\
\rho^{-B}
\end{array}
\right)
,\quad\eta=\left(
\begin{array}{c}
\eta^0\\
\eta^-\\
\eta^{-A}
\end{array}
\right)
\end{equation}\label{eq:charges}
where each triplet belongs to $(1,3,X)$ with
\be
X_\chi=\beta_{Q}/\sqrt3,\quad X_\rho=1/2-\beta_{Q}/(2\sqrt3),\quad X_\eta=-1/2-\beta_{Q}/(2\sqrt3).
\ee

In addition to neutral and singly charged states, there are fields with charge
\begin{equation}\label{eq:ABcharges}
Q^A=\frac{1}{2}+\frac{\sqrt3}{2}\beta_{Q}\;,\quad Q^B=-\frac{1}{2}+\frac{\sqrt3}{2}\beta_{Q}.
\end{equation}

The symmetry breaking pattern $SU(3)_L\times U(1)_X\to U(1)_{Q}$ is obtained in two steps. Firstly, the vacuum expectation value (VEV) of the neutral component of $\chi$ gives masses to the extra gauge bosons and the extra quarks, triggering the pattern $SU(3)_L\times U(1)_X\to SU(2)_L\times U(1)_Y$. Afterwards, the usual spontaneous symmetry breaking mechanism $SU(2)_L\times U(1)_Y\to U(1)_{Q}$ is realised by the VEVs of the neutral components of $\rho$ and $\eta$.

The scalar potential reads 
\begin{align}\label{pot}
V&= m^2_1\, \rho^*\rho+m^2_2\,\eta^*\eta+m^2_3\,\chi^*\chi +\sqrt2 f_{\rho\eta\chi} \rho\, \eta\, \chi\nn\\
&+\lambda_1 (\rho^*\rho)^2+\lambda_2(\eta^*\eta)^2+\lambda_3(\chi^*\chi)^2\nn\\
&+\lambda_{12}\rho^*\rho\,\eta^*\eta+\lambda_{13}\rho^*\rho\,\chi^*\chi+\lambda_{23}\eta^*\eta\,\chi^*\chi\nn\\
&+\zeta_{12}\rho^*\eta\,\eta^*\rho+\zeta_{13}\rho^*\chi\,\chi^*\rho+\zeta_{23}\eta^*\chi\,\chi^*\eta,
\end{align}
where the neutral component of each triplet is expanded around its VEV as
\begin{align}
\rho^0
&
\to\frac{1}{\sqrt2}\left(v_\rho+\rm{Re}\,\rho^0+i\,\rm{Im}\,\rho^0\right),\\
\eta^0
&
\to\frac{1}{\sqrt2}\left(v_\eta+\rm{Re}\,\eta^0+i\,\rm{Im}\,\eta^0\right),\\
\chi^0
&
\to\frac{1}{\sqrt2}\left(v_\chi+\rm{Re}\,\chi^0+i\,\rm{Im}\,\chi^0\right).
\end{align}
The dimensionless parameter $\kappa$, that stems from the redefinition $f_{\rho\eta\chi}=\kappa v_\chi$, can be also introduced to deal with only two scales, $v_\chi$ and $v=\sqrt{v_\eta^2+v_\rho^2}$. The ratio of the two light VEVs is conveniently redefined via $\tan \beta=v_\eta/v_\rho$.\\

The minimisation conditions of the potential, defined by $\left.\partial V/\partial\Phi\right|_{\Phi=0}=0$, are given by
\bea
m^2_1v_\rho+\lambda_1v_\rho^3+\frac{\lambda_{12}}{2}v_\rho v_\eta^2-f_{\rho\eta\chi}v_\eta v_\chi+\frac{\lambda_{13}}{2}v_\rho v_\chi^2=0,\\
m^2_2v_\eta+\lambda_2v_\eta^3+\frac{\lambda_{12}}{2}v_\rho^2 v_\eta-f_{\rho\eta\chi}v_\rho v_\chi+\frac{\lambda_{23}}{2}v_\eta v_\chi^2=0,\\
m^2_3v_\chi+\lambda_3 v_\chi^3+\frac{\lambda_{13}}{2}v_\rho^2 v_\chi-f_{\rho\eta\chi}v_\rho v_\eta+\frac{\lambda_{23}}{2}v_\eta^2 v_\chi=0.
\eea

Using these conditions, one can compute the mass matrices for the scalar degrees of freedom. In~\ref{sec:diag}, a diagonalisation procedure is discussed in details. Moreover, it is shown how the ten parameters of the scalar potential $\lambda_1$, $\lambda_2$, $\lambda_3$, $\lambda_{12}$, $\lambda_{13}$, $\lambda_{23}$, $\zeta_{12}$, $\zeta_{13}$, $\zeta_{23}$, $\kappa$, can be traded for the three physical neutral masses $m_{h_1}$, $m_{h_2}$, $m_{h_3}$, the singly charged mass $m_{h_1^\pm}^2$, the two $A$- and $B$-charged masses $m^2_{h_1^{\pm A}}$, $m^2_{h_1^{\pm B}}$, the pseudoscalar mass $m^2_{a_1}$ and the three mixing angles $\alpha_1$, $\alpha_2$ and $\alpha_3$ of the neutral scalar fields. This is considerably important to allow for an automation in benchmark selections for future phenomenological analysis.

A final remark is required before discussing the following sections. The cases $\beta_{Q}=\pm1/\sqrt3$ support an extended potential with extra terms among fields. This occurs because two of the scalars that generally belong to different representations would eventually collapse into the same representation of $SU(3)_L\times U(1)_X$. The potential can be protected from developing these terms by imposing a new extra symmetry. We report as an example a global $U(1)$ symmetry acting on the potential (\emph{à la} Froggatt--Nielsen~\cite{Huitu:2017ukq}). Still, an extra set of two neutral scalars could support the presence of a fourth VEV\footnote{Formally, one should also introduce a fifth VEV that would always be reabsorbed by a suitable transformation of the fields.} that triggers the first symmetry breaking pattern together with $v_\chi$. If the extra $U(1)$ symmetry is kept intact the minimisation conditions would call for a zero value of the new VEV, thus restoring the setup where only three VEVs are included. However, such a model will come along with specific phenomenological implications. In fact, the exact global $U(1)$ symmetry will force a pseudoscalar to be massless. On the contrary, when the symmetry is softly broken, the physical spectrum changes and our conclusions do not hold anymore. For these reasons, the study of the scalar potential in the $\beta_{Q}=\pm1/\sqrt3$ case requires a dedicated analysis, far beyond the scope of this work.

\section{Boundedness from below}\label{sec:bfb}
\noindent
In models with many Higgs fields, proving the existence of a finite absolute minimum for the scalar potential is generally a delicate issue.
\ Such a matter presents many issues that have been often addressed~\cite{Hadeler1983,Klimenko:1984qx,Ivanov:2018jmz}, especially in models with many Higgs doublets~\cite{Maniatis:2006fs,Degee:2012sk,Kannike:2012pe,Maniatis:2015gma,Kannike:2016fmd,Faro:2019vcd}. In this section, a solution for the general $331$ Model is delivered: a set of necessary and sufficient conditions to ensure the BFB of the potential in Eq.~\ref{pot} is presented.

In general, one has to analyse the behaviour of the highest powers of the fields, \emph{i.e.} the properties of the quartic couplings of the ultraviolet-complete theory. For this purpose, it is convenient to parameterise a triplet in the following form:
\begin{equation}
\Phi_i = \sqrt{r_i} e^{i\,\gamma_i}\left(
\begin{array}{c}
\sin a_i \cos b_i\\
e^{i\,\beta_i} \sin a_i \sin b_i\\
e^{i\,\alpha_i}\cos a_i
\end{array}
\right),\qquad i=1,2,3,
\end{equation}
where the fields are complex numbers, $a_i,\,b_i,\,\beta_i,\,\gamma_i$ and $\alpha_i$ are angular parameters, and $r_i$ is the radial part of the field. Manifestly, one finds $\Phi_i^\dagger \Phi^{\phantom{\dagger}}_i = r_i \geq 0$.

For the sake of convenience, the following quantity is introduced:
\begin{align}\label{zijgen}
\tau_{ij} & = \left(\Phi_i^\dagger \Phi^{\phantom{\dagger}}_i\right)\, \left(\Phi_j^\dagger \Phi^{\phantom{\dagger}}_j\right) - \left(\Phi_i ^\dagger \Phi^{\phantom{\dagger}}_j\right)\, \left(\Phi_j^\dagger \Phi^{\phantom{\dagger}}_i\right),
\end{align}
where the non-negativity of $\tau_{ij}$ is ensured by the Cauchy--Schwarz inequality.

Then, the underlying gauge symmetry allows to write the triplet fields in the form
\begin{equation}\label{fieldgauge}
\frac{\Phi_1}{\sqrt{r_1}}=\left(
\begin{array}{c}
0\\
0\\
1
\end{array}
\right),
\frac{\Phi_2}{\sqrt{r_2}}=\left(
\begin{array}{c}
0\\
\sin a_2\\
e^{i\alpha_2}\cos a_2
\end{array}
\right),
\frac{\Phi_3}{\sqrt{r_3}}=\left(
\begin{array}{c}
\sin a_3 \cos b_3\\
\sin a_3 \sin b_3\\
e^{i\alpha_3} \cos a_3
\end{array}
\right).
\end{equation}

The quartic part of the scalar potential in Eq.~\ref{pot} can be written 
in terms of 
a \textit{radial} and an \textit{angular} block:
\begin{align}
V^{(4)} &= V_R + \zeta^\prime_{12}\tau^{\phantom{\prime}}_{12}+\zeta^\prime_{13}\tau^{\phantom{\prime}}_{13}+\zeta^\prime_{23}\tau^{\phantom{\prime}}_{23} = V_R + V_A,
\end{align} 
where the $\zeta$ parameter were conveniently traded with $\zeta^\prime_{ij} = - \zeta^{\phantom{\prime}}_{ij}$ and the radial part reads
\begin{align}
V_R&=\lambda_1 (\rho^*\rho)^2+\lambda_2(\eta^*\eta)^2+\lambda_3(\chi^*\chi)^2\nn\\
&+\lambda^\prime_{12}\rho^*\rho\,\eta^*\eta+\lambda^\prime_{13}\rho^*\rho\,\chi^*\chi+\lambda^\prime_{23}\eta^*\eta\,\chi^*\chi,
\end{align}
with $\lambda^\prime_{ij} = \lambda^{\phantom{\prime}}_{ij}+\zeta^{\phantom{\prime}}_{ij}$.

The radial part of the scalar potential has no dependence on the angular parameters of the fields. Conversely, $V_{A}=\zeta^\prime_{12}\tau^{\phantom{\prime}}_{12}+\zeta^\prime_{13}\tau^{\phantom{\prime}}_{13}+\zeta^\prime_{23}\tau^{\phantom{\prime}}_{23}$ depends on both radial and angular variables. The two blocks should be analysed separately.

The BFB of the radial part of the scalar potential is obtained by imposing the co-positivity constraints \cite{Hadeler1983,Kannike:2012pe,Faro:2019vcd} on the matrix $Q_{ij}$, defined by
\begin{equation}
V_R \equiv Q_{ij}r_i r_j.
\end{equation}
This is the set of necessary and sufficient conditions for the BFB of the potential for the case $\zeta^\prime_{12}=\zeta^\prime_{13}=\zeta^\prime_{23}=0$.

A good strategy to get rid of the angular information of $V_{A}$ is to search for the ``angular minima''. This can partially solve the problem and give an ``angularly minimised'' scalar potential with radial dependence only. On top of this, one can apply the co-positivity criterion on $V_R+V_{A}$, thus obtaining a complete set of necessary and sufficient conditions~\cite{Faro:2019vcd}.

Using the parameterisation in Eqs.~\ref{fieldgauge}, one can firstly minimise $V_A$ along the phase direction by imposing $\partial_\delta V_{A}=0$ with $\delta\equiv\alpha_{2}-\alpha_{3}$, and secondly obtain the following (normalised) components of the $V_A$ gradient:
\begin{align}\label{minpotcb}
\frac{\partial_{a_2}V_{A}}{r_2}&=\sin 2a^{\phantom{\prime}}_2\, r^{\phantom{\prime}}_1 \, \zeta^\prime_{12}\nn\\
&+(\sin 2a^{\phantom{\prime}}_3\cos 2a^{\phantom{\prime}}_2 \sin b^{\phantom{\prime}}_3 \nn\\
&+ \sin{2a^{\phantom{\prime}}_2}(\cos^2{a^{\phantom{\prime}}_3}-\sin^2a^{\phantom{\prime}}_3\sin^2b^{\phantom{\prime}}_3))\, r^{\phantom{\prime}}_3\, \zeta^\prime_{23},\nn\\
\frac{\partial_{a_3}V_{A}}{r_3}&=\sin 2a^{\phantom{\prime}}_3\, r^{\phantom{\prime}}_1 \, \zeta^\prime_{13}\\
&+(\sin 2a^{\phantom{\prime}}_2\cos 2a^{\phantom{\prime}}_3 \sin b^{\phantom{\prime}}_3 \nn\\
&+ \sin 2a^{\phantom{\prime}}_3(\cos^2a^{\phantom{\prime}}_2-\sin^2a^{\phantom{\prime}}_2\sin^2b^{\phantom{\prime}}_3))\,r^{\phantom{\prime}}_2 \, \zeta^\prime_{23},\nn\\
\frac{\partial_{b_3}V_{A}}{r_2 r_3}&=\frac{1}{2}\cos b^{\phantom{\prime}}_3(\sin 2a^{\phantom{\prime}}_2 \sin 2a^{\phantom{\prime}}_3-4\sin^2a^{\phantom{\prime}}_2\sin^2a^{\phantom{\prime}}_3\sin b^{\phantom{\prime}}_3)\,\zeta^\prime_{23}.\nn
\end{align}
Setting all the components of the $V_A$ gradients equal to zero and solving the correspondent system of equations lead to trivial solutions for $a_2,a_3,b_3 = k\, \pi/2$ with $k\in\mathbb Z$. Correspondingly, one obtains four angular minima for $V_{A}$:
\begin{align}
 \mbox{min}(V_A)^{T}_{1}&=\zeta^\prime_{12}\,r^{\phantom{\prime}}_1 r^{\phantom{\prime}}_2 + \zeta^\prime_{23}\,r^{\phantom{\prime}}_2 r^{\phantom{\prime}}_3, \\
 \mbox{min}(V_A)^{T}_{2}&=\zeta^\prime_{13}\,r^{\phantom{\prime}}_1 r^{\phantom{\prime}}_3 + \zeta^\prime_{23}\,r^{\phantom{\prime}}_2 r^{\phantom{\prime}}_3, \\ 
 \mbox{min}(V_A)^{T}_{3}&=\zeta^\prime_{12}\, r^{\phantom{\prime}}_1 r^{\phantom{\prime}}_2 + \zeta^\prime_{13}\, r^{\phantom{\prime}}_1 r^{\phantom{\prime}}_3,\\
 \mbox{min}(V_A)^{T}_{4}&=\zeta^\prime_{12}\, r^{\phantom{\prime}}_1 r^{\phantom{\prime}}_2 + \zeta^\prime_{13}\, r^{\phantom{\prime}}_1 r^{\phantom{\prime}}_3 + \zeta^\prime_{23}\, r^{\phantom{\prime}}_2 r^{\phantom{\prime}}_3.
\end{align}

After the minimisation of the angular part of the potential, the BFB conditions call for co-positivity constraints applied on the new matrices $\widetilde{Q}_k$ defined by
\begin{equation}
V_{R} + \mbox{min}(V_{A})^{T}_{k} = \widetilde{Q}_{k}^{ij} r^{\phantom{\prime}}_i r^{\phantom{\prime}}_j,\,\quad k=1,\ldots,4.
\end{equation}
This is the set of necessary and sufficient conditions for the BFB of the potential when at least one of the $\zeta^\prime$ is zero.

Apart from the trivial stable points described above, there can be more solutions 
 to the system of equations $\partial_{i} V_{A}=0$ with $i=a_2,a_3,b_3$. 
For the convenience of the reader, they can be written in the following form:
\begin{align}
\partial^{\phantom{\prime}}_{a_2}V_{A} &= f(a^{\phantom{\prime}}_2)\, r^{\phantom{\prime}}_1 r^{\phantom{\prime}}_2\, \zeta^\prime_{12} + g(a^{\phantom{\prime}}_2,a^{\phantom{\prime}}_3,b^{\phantom{\prime}}_3)\, r^{\phantom{\prime}}_2 r^{\phantom{\prime}}_3\, \zeta^\prime_{23},\\
\partial^{\phantom{\prime}}_{a_3}V_{A} &= f(a^{\phantom{\prime}}_3)\, r^{\phantom{\prime}}_1 r^{\phantom{\prime}}_3\, \zeta^\prime_{13} + h(a^{\phantom{\prime}}_2,a^{\phantom{\prime}}_3,b^{\phantom{\prime}}_3)\, r^{\phantom{\prime}}_2 r^{\phantom{\prime}}_3\, \zeta^\prime_{23},\\ \label{eq:db3va}
\partial^{\phantom{\prime}}_{b_3}V_{A} &= k(a^{\phantom{\prime}}_2,a^{\phantom{\prime}}_3,b^{\phantom{\prime}}_3)\, r^{\phantom{\prime}}_2 r^{\phantom{\prime}}_3\, \zeta^\prime_{23}.
\end{align}

Since Eq.~\ref{eq:db3va} does not contain any radial information, the best strategy to search for more stable points is to set it equal to zero and find solutions for all the angular variables. It can be proven that all the solutions obtained with respect to $a_2$ and $a_3$ lead again to the trivial cases discussed above. On the other hand, when 
 
$b_3$ is considered, one finds candidates for $\tilde b_3 \equiv b_3(a_2,a_3)$ that lead to
\begin{equation}
g(a_2,a_3,\tilde b_3)=h(a_2,a_3,\tilde b_3).
\end{equation}
Therefore, the requirement $\partial_{i} V_{A}=0$ with $i=a_2,a_3,\tilde b_3$ implies that
\begin{equation}\label{ivan}
f(a^{\phantom{\prime}}_2)\, r^{\phantom{\prime}}_1 r^{\phantom{\prime}}_2\, \zeta^\prime_{12}= - g(a^{\phantom{\prime}}_2,a^{\phantom{\prime}}_3,\tilde b^{\phantom{\prime}}_3)\, r^{\phantom{\prime}}_2r^{\phantom{\prime}}_3\, \zeta^\prime_{23} = f(a^{\phantom{\prime}}_3)\, r^{\phantom{\prime}}_1r^{\phantom{\prime}}_3\,\zeta^\prime_{13}.
\end{equation}
The explicit solutions are $f(x)=\sin 2x$ and $g(a_2,a_3,\tilde b_3)= \sin(a_2-a_3)$,
\ which are the only non-trivial stable points of $V_{A}$.

By analogy with the analysis in~\cite{Faro:2019vcd}, Eq.~\ref{ivan} can be interpreted as the Law of Sines of a triangle. In such framework, it is proven that the non-trivial cases can be recast in the following expression:
\begin{equation}
\mbox{min}(V_{A})^{NT} = \frac{\zeta^\prime_{12}\zeta^\prime_{13}\zeta^\prime_{23}}{4}\left(\frac{r_1}{\zeta^\prime_{23}}+\frac{r_2}{\zeta^\prime_{13}}+\frac{r_3}{\zeta^\prime_{12}}\right)^2.
\end{equation}  
In order to grant the BFB of the scalar potential, the matrix $\widehat{Q}_{ij}$ defined by
\begin{equation}
V_{R} + \mbox{min}(V_{A})^{NT} = \widehat{Q}^{ij} r_i r_j,
\end{equation}
is also required to fulfil the co-positivity criterion, once the transformation described in~\cite{Faro:2019vcd} is applied. Hence, the co-positivity of \emph{both} $\widetilde{Q}$ \emph{and the transformed} $\widehat{Q}$ is the necessary and sufficient condition to have a stable potential when all the $\zeta^\prime$ are different from zero.
The co-positivity criteria for a generic rank-3 matrix $A$ are
\begin{align}\label{coposit1}
&A_{ii} \geq 0,\quad \mbox{with }i=1,2,3,
\\
& \mathring{A}_{ij}\equiv \sqrt{A_{ii}A_{jj}} + A_{ij} \geq 0,\quad \mbox{with }i,j=1,2,3,
\\
&\sqrt{A_{11}A_{22}A_{33}} + A_{12}\sqrt{A_{33}} + A_{13}\sqrt{A_{22}} + A_{23}\sqrt{A_{11}}\nn\\
&+ \sqrt{2\mathring{A}_{12}\mathring{A}_{13}\mathring{A}_{23}} \geq 0\,. \label{copositivity2}
\end{align}
These are the first conditions implemented in the Mathematica file described in~Section~\ref{sec:intro}.

\section{Perturbative Unitarity}\label{sec:unitarity}
\noindent
The methodology to obtain perturbative unitarity constraints on the SM was described for the first time in~\cite{Lee:1977eg}:
\ all the possible $2 \rightarrow 2$ processes with a given total charge $\mathcal{Q}$ should be considered and the corresponding amplitudes arranged in a scattering matrix. Each row (column) of this matrix corresponds to a different possible initial (final) state. For each element of this matrix, from the partial-wave expansion of the corresponding amplitude $\mathcal{A}(s,\theta)$, only the spherical wave
\begin{equation}
 a_0 = \frac{1}{32\pi} \int_{-1}^1 d\cos\theta \mathcal{A}(s,\theta)
\end{equation}
should be retained, as it is known to give the
\ leading contribution at large energies. The perturbative unitarity condition imposes then that the real part of the largest eigenvalue of this matrix should not exceed $1/2$. Because what matters is the behaviour at large energies, the calculation can be simplified by replacing the external vector bosons with the corresponding Goldstone bosons, in accordance with the equivalence theorem~\cite{Lee:1977eg,Chanowitz:1985hj,Gounaris:1986cr}. 

In the general $331$ Model, the scalars of Eq.~\ref{eq:scalars} can have charges $0$, $\pm1$, $\pm Q^A$ and $\pm Q^B$, where $Q^A$ and $Q^B$ take different values depending on the specific realisation of the $331$ Model, \textit{i.e.} of the value of $\beta_{Q}$ (see Eq.~\ref{eq:ABcharges}). It follows that there are $13$ scattering matrices, corresponding to the initial total charge of the $2 \rightarrow 2$ processes 
\begin{align}\label{eq:charge_scattering_matrices}
\mathcal{Q}&=0,\,1,\,2,\,Q^A,\,Q^B,\,Q^A+1,\,Q^B+1,\,Q^A-1,\,Q^B-1,\, \nonumber \\
&\quad Q^A+Q^B,\,Q^A-Q^B,\,2Q^A,\,2Q^B.
\end{align}
Notice that opposite charge signs would lead to equivalent constraints. The final condition for the perturbative unitarity is then
\begin{equation}
 |\mathbf{a}|\leq\frac{1}{2}
\end{equation}
where $\mathbf{a}$ identifies all eigenvalues. Their form is shown in the following list
\begin{align}
\mathbf{a}=\Big\{&\frac{\lambda_i}{8\pi},\frac{\lambda_{ij}}{16\pi},\frac{\lambda_{ij}\pm\zeta_{ij}}{16\pi},\frac{\lambda_{ij}+2\zeta_{ij}}{16\pi},\frac{\lambda_i+\lambda_j\pm\sqrt{(\lambda_i - \lambda_j)^2 + \zeta_{ij}^2}}{16\pi},\nn\\
 &\frac{\mathcal{P}_1^3 (\lambda_m,\lambda_{mn},\zeta_{mn})}{32\pi},\frac{\mathcal{P}_2^3 (\lambda_m,\lambda_{mn},\zeta_{mn})}{32\pi}\Big\}
\end{align}
where $\mathcal{P}_{1,2}^3$ are the solutions of third-grade polynomials given by

\begin{align}
& \sum_{i,j,k=1}^3\left[ \frac{x^3}{27}-\frac{4}{9} \lambda_ix^2
+\Big(2 \left( 4 \lambda_i \lambda_j-\zeta_{ij}^2\right)x-\frac{8}{3}\big(\zeta_{ij} \zeta_{ik} \zeta_{jk}-3 \lambda_i \zeta_{jk}^2\right.\nn\\
&\left.\qquad\qquad\qquad\qquad\qquad\qquad\qquad\quad+4 \lambda_i \lambda_j \lambda_k\big)\Big)\left(\varepsilon_{ijk}\right)^2\right],\\
 &\sum_{i,j,k=1}^3\left[ \frac{x^3}{27}
-\frac{16}{9} \lambda_ix^2
+\Big(2 ( 64 \lambda_i \lambda_j-(3 \lambda_{ij}+\zeta_{ij})^2)x\nn \right.\\
&\left.\qquad\qquad-\frac{8}{3}\big(
\zeta_{ik} \zeta_{jk} (9 \lambda_{ij}+\zeta_{ij})+27 \lambda_{ij} \lambda_{ik} (\lambda_{jk}+\zeta_{jk})\right.\nn\\
&\left.\qquad\qquad+4 \lambda_{i} \big(64 \lambda_{j} \lambda_{k}-3 (3 \lambda_{jk}+\zeta_{jk})^2\big)
\big)\Big)\left(\varepsilon_{ijk}\right)^2\right],
\end{align}
with $\lambda_{ji}=\lambda_{ij}$, $\zeta_{ji}=\zeta_{ij} $.

It is easy to see that this condition does not change when $Q^A$ and $Q^B$ are such that some values of $\mathcal{Q}$ in Eq.~\ref{eq:charge_scattering_matrices} are equal: the total number of matrices would be smaller than $13$, and the matrices corresponding to the equal values of $\mathcal{Q}$ would have blocks, making the condition on the largest eigenvalue unchanged. Therefore, the method outlined above is valid for the general $331$ Model, without any need to specify the value of $\beta_Q$. The only exception is the case $\beta=\pm 1/\sqrt{3}$, where the Lagrangian can contain more possible interactions among the scalars and the rank of the scattering matrices can be larger. This case must be treated separately, unless a global $U(1)$ symmetry is imposed. Such a symmetry would bring the scalar potential back to Eq.~\ref{pot}, as discussed in Section~\ref{sec:model}, and the calculation of the perturbative unitarity constraint with generic $\beta_{Q}$ would then apply again.

Given the dimensionality of the scattering matrices, the diagonalisation problem should be solved numerically due to the presence of the third-degree polynomials. 
For this purpose, a Mathematica file was developed to perform the computation described above: a FeynRules model file was created and the FeynArts interface of FeynRules was exploited to produce a model file for the FeynArts and FormCalc packages; furthermore, the combined packages FeynArts and FormCalc were linked to the aforementioned Mathematica file to allow for an automated numerical approach to the study of perturbative unitarity constraints in the general $331$ Model.

\section{Perturbativity}\label{sec:perturbativity}
\noindent
Requesting perturbative unitarity to be respected is necessary but not sufficient to guarantee the correct perturbative behaviour of the model. Perturbativity of the couplings should also be enforced,
\ setting further theoretical constraints on the parameters of the model.

Perturbativity constraints act on the adimensional couplings according to the condition $|\lambda_J|\leq 4\pi k$, where $k\leq 1$ is an arbitrary parameter designed to tune the bound.

These constraints turn out to be especially effective when the couplings of the scalar potential are recast in terms of physical parameters according to the diagonalisation procedure described in~\ref{sec:diag}.

Even if the explicit expressions are too cumbersome and difficult to interpret, it is always possible to consider a limiting case that comprise some phenomenological information. In fact, the general $331$ Model is built upon a gauge symmetry that is larger than the SM ones. This implies that VEV responsible for the $SU(3)\times U(1)\to SU(2)\times U(1)$ symmetry-breaking pattern has to be (much) larger than the electroweak scale. Consequently, one should consider the limit $v_\chi\gg v$
\ and expand the explicit expressions for the adimensional couplings given in Eqs.~\ref{parsoll1}-\ref{parsolkap} up to the first meaningful order:
\begin{align}
\lambda_1&= -\frac{m^2_{a_1} \tan ^2\beta}{2 v^2}+m^2_{h_1} \frac{\mathtt{C}^2_2 \mathtt{C}^2_3 \sec ^2\beta}{2 v^2}\nn\\
   &+m^2_{h_2}\frac{ \sec ^2\beta (\mathtt{S}_1
   \mathtt{S}_2 \mathtt{C}_3-\mathtt{C}_1 \mathtt{S}_3)^2}{2 v^2}\nn\\
   &+m^2_{h_3}\frac{ \sec
   ^2\beta (\mathtt{C}_1 \mathtt{S}_2 \mathtt{C}_3+\mathtt{S}_1 \mathtt{S}_3)^2}{2
   v^2}+O\left(\frac{m}{v_\chi}\right) \label{eq:lamexp1} \\
\lambda_2&=-\frac{m^2_{a_1} \cot ^2\beta}{2 v^2}+m^2_{h_1}\frac{ \mathtt{C}^2_2 \mathtt{S}^2_3 \csc ^2\beta}{2 v^2}\nn\\
    &+m^2_{h_2}\frac{ \csc ^2\beta \left(\mathtt{S}^2_1 \mathtt{S}^2_2 \mathtt{S}^2_3+2\mathtt{S}_1 \mathtt{C}_1 \mathtt{S}_2 \mathtt{C}_3\mathtt{S}_3+\mathtt{C}^2_1 \mathtt{C}^2_3\right)}{2
   v^2}\nn\\
   &+m^2_{h_3}\frac{ \csc ^2\beta \left(\mathtt{C}_1 \mathtt{S}_2 \left(\mathtt{C}_1 \mathtt{S}_2 \mathtt{S}^2_3-2\mathtt{S}_1 \mathtt{C}_3\mathtt{S}_3\right)+\mathtt{S}^2_1 \mathtt{C}^2_3\right)}{2
   v^2}+O\left(\frac{m}{v_\chi}\right), \label{eq:lamexp2}
   \end{align}
   \begin{align}
\lambda_{12}&=\frac{m^2_{a_1}}{v^2}+m^2_{h_1}\frac{ \mathtt{C}^2_2 \mathtt{S}_3 \mathtt{C}_3 \csc \beta \sec
   \beta}{v^2}\nn\\
   &+m^2_{h_2}\frac{ \csc \beta \sec \beta}{4 v^2}
   \left(4 \mathtt{C}_1\mathtt{S}_1 \mathtt{S}_2(\mathtt{C}_3^2-\mathtt{S}_3^2) \right.\nn\\
   &\left.-\mathtt{C}_3\mathtt{S}_3 \left(2 \mathtt{S}^2_1(\mathtt{C}_2^2-\mathtt{S}_2^2) +6 \mathtt{C}_1\mathtt{S}_1+1\right)\right)\nn\\
   &-m^2_{h_3}\frac{ \csc \beta \sec \beta}{4
   v^2}
   \left(4 \mathtt{C}_1\mathtt{S}_1
   \mathtt{S}_2(\mathtt{C}_3^2-\mathtt{S}_3^2)\right.\nn\\
   &\left.+\mathtt{C}_3\mathtt{S}_3 \left(2 \mathtt{C}^2_1(\mathtt{C}_2^2-\mathtt{S}_2^2) -3(\mathtt{C}_1^2-\mathtt{S}_1^2)+1\right)\right)+O\left(\frac{m}{v_\chi}\right),\\
\zeta_{12}&=\frac{2}{v^2}\left(m^2_{h_1^\pm}-m^2_{a_1^{\phantom{\pm}}}\right)+O\left(\frac{m}{v_\chi}\right),\label{eq:zetaexp}\\
\lambda_3&=\lambda_{13}=\lambda_{23}=\zeta_{13}=\zeta_{23}=O\left(\frac{m}{v_\chi}\right).\label{eq:lamexpn}
\end{align}
Ideally, all the terms in the right-hand sides of the Eqs.~\ref{eq:lamexp1}-\ref{eq:lamexpn} should combine to keep the couplings on the left-hand sides below the perturbativity threshold. Remarkably, for a mass spectrum that lives above the electroweak VEV, Eq.~\ref{eq:zetaexp} calls for a certain degree of degeneracy between $m^2_{h_1^\pm}$ and $m^2_{a_1^{\phantom{\pm}}}$. Beyond that, the set of Eqs.~\ref{eq:lamexp1}-\ref{eq:lamexpn} does not provide any general take-home messages. Even if specific benchmark choices could lead to more manageable formulae, a numerical approach is always required to investigate generic scenarios. Consequently, the perturbativity conditions on the left-handed side of Eqs.~\ref{parsoll1}-\ref{parsolkap} were unified with the bounds presented in Sections~\ref{sec:bfb}~and~\ref{sec:unitarity} to complete the Mathematica file designed for a fast selection of theoretically allowed portions of the general $331$ Model parameter space.

\section{Conclusions}\label{sec:conclusions}
\noindent
Theoretical constraints play a key role in selecting the viable portion of the parameter space of multi-Higgs models. Specifically, any scalar potential has to fulfil the requirements of boundedness from below and perturbativity of the couplings. Moreover, scattering matrices must satisfy perturbative unitarity conditions. In this article these constraints were studied in the context of the general $331$ Model. 

For the first time, these constraints were systematically analysed and combined in a framework that allows for fast numerical checks of specific $331$ Model benchmarks.

The present analysis of the boundedness from below of the scalar potential led to a set of necessary and sufficient conditions specific to the general $331$ Model that were overlooked in previous literature.
\ As a general approach of this work, the Lagrangian parameters were expressed in terms of the physical parameters, namely masses and mixing angles, by means of a systematic diagonalisation of all the mass matrices of the scalar sector. Perturbativity and perturbative unitarity were then discussed in this spirit and maintaining a consistent general approach.

All these results were added together in a Mathematica file to open the way for future collider studies of specific realisations of the $331$ Model in light of a systematic analysis of the parameter space of the scalar sector.

\section*{Acknowledgements}\label{sec:ack}
\noindent
AC acknowledges Igor Ivanov for pointing out the issue of the boundedness from below of the scalar potential in multi-Higgs models and for useful discussions on this issue. The Authors acknowledge Gennaro Corcella and Luca Panizzi for collaborating during the early stage of the project and for many useful discussion throughout its development.

\appendix

\section{Diagonalisation of the Scalar Sector}\label{sec:diag}
\noindent
After the spontaneous symmetry breaking mechanism, the gauge eigenstates $\rho$, $\eta$ and $\chi$ are rotated into the 
mass eigenstates. First of all, the parameter $\kappa$ is directly related to the massive pseudoscalar state obtained from the CP-odd neutral scalars. In fact, with 
\be
a_i=\mathcal{R}^P_{ij} A_j,
\ee
where $\vec A=(\rm{Im}\,\rho^0,\,\rm{Im}\,\eta^0,\,\rm{Im}\,\chi^0)$ and $\vec a=(a_{G_Z},\,a_{G_{Z^{'}}},\,a_1)$, the mass matrix\footnote{The mass matrices of the scalar sector are expressed in the unitary gauge, where the Nambu--Goldstone bosons are the eigenvectors corresponding to the null eigenvalues.} for the neutral pseudoscalars is

\begin{equation}
\mathcal{M}_a^2=\left(
\begin{array}{ccc}
 \kappa v_\chi^2 \tan \beta & \kappa v_\chi^2 & \kappa v_\chi v \sin \beta \\
  \sim& \kappa v_\chi^2 \cot \beta & \kappa v_\chi v \cos \beta \\
 \sim&\sim & \kappa v^2
   \cos \beta \sin \beta \\
\end{array}
\right),
\end{equation}
hence
\begin{equation}\label{pseudomass}
\kappa =\frac{m^2_{a_1}}{(v_\chi^2 \csc \beta \sec \beta + v^2 \cos \beta \sin \beta) }.
\end{equation} 

In the CP-even neutral sector, the rotation implies
\begin{equation}
h_i=\mathcal{R}^S_{ij} H_j
\end{equation}
where $\vec H=(\rm{Re}\,\rho^0,\,\rm{Re}\,\eta^0,\,\rm{Re}\,\chi^0)$ and $\vec h=(h_1,\,h_2,\,h_3)$. The explicit expression of the mass matrix of the neutral scalars is given by
\begin{align}
\mathcal{M}^2_{h;1,1}&= \kappa \tan \beta v_\chi^2+2 \lambda_1 v^2 \cos ^2\beta, \\
\mathcal{M}^2_{h;2,2}&= \kappa \cot \beta v_\chi^2+2\lambda_2 v^2 \sin ^2\beta,\\
\mathcal{M}^2_{h;3,3}&=2 \lambda_3 v_\chi^2+\kappa v^2 \cos \beta \sin \beta,\\
\mathcal{
M}^2_{h;1,2}&= \lambda_{12} v^2 \cos \beta\sin \beta-\kappa v_\chi^2, \\
\mathcal{
M}^2_{h;1,3}&=v_\chi v (\lambda_{13} \cos \beta-\kappa \sin \beta), \\
\mathcal{
M}^2_{h;2,3}&=v_\chi v (\lambda_{23} \sin \beta-\kappa \cos \beta).\label{mmh2}
\end{align}

Therefore, the diagonalisation is realised by means of an orthogonal rotation involving three different mixing angles:
\begin{equation}\label{3drot}
\mathcal{R}^S=\left(
\begin{array}{ccc}
 \mathtt{C}_2 \mathtt{C}_3 & \mathtt{C}_3 \mathtt{S}_1 \mathtt{S}_2-\mathtt{C}_1 \mathtt{S}_3 & \mathtt{C}_1 \mathtt{C}_3 \mathtt{S}_2+\mathtt{S}_1 \mathtt{S}_3 \\
 \mathtt{C}_2 \mathtt{S}_3 & \mathtt{C}_1 \mathtt{C}_3+\mathtt{S}_1 \mathtt{S}_2 \mathtt{S}_3 & \mathtt{C}_1 \mathtt{S}_2 \mathtt{S}_3-\mathtt{C}_3 \mathtt{S}_1 \\
 -\mathtt{S}_2 & \mathtt{C}_2 \mathtt{S}_1 & \mathtt{C}_1 \mathtt{C}_2 \\
\end{array}
\right)
\end{equation} 
where $\mathtt{C}_i\equiv \cos\alpha_i$ and $\mathtt{S}_i\equiv \sin\alpha_i$.
\ This leads to the physical scalar mass matrix
\begin{equation}\label{eig}
\mathcal{\hat M}^2_h=\mathtt{diag}(m_{h_1}^2,m_{h_2}^2,m_{h_3}^2)=(\mathcal{R}^S)^T\cdot\mathcal{M}_h^2\cdot\mathcal{R}^S.
\end{equation}

The solution of the diagonalisation conditions given in Eqs.~\ref{pseudomass}~and~\ref{eig} is

\begin{align}
\lambda_1&=-m^2_{a_1}\frac{4
    v_\chi^2 \tan ^2\beta}{8 v_\chi^2 v^2+v^2(1-\cos4\beta)}+m^2_{h_1} \frac{\mathtt{C}^2_2 \mathtt{C}^2_3 \sec ^2\beta}{2 v^2}\label{parsoll1}\nn\\
   &+m^2_{h_2}\frac{ \sec ^2\beta (\mathtt{S}_1
   \mathtt{S}_2 \mathtt{C}_3-\mathtt{C}_1 \mathtt{S}_3)^2}{2 v^2}\nn\\
   &+m^2_{h_3}\frac{ \sec
   ^2\beta (\mathtt{C}_1 \mathtt{S}_2 \mathtt{C}_3+\mathtt{S}_1 \mathtt{S}_3)^2}{2
   v^2},
\end{align}
\begin{align}    
\lambda_2&=-m^2_{a_1}\frac{4
    v_\chi^2 \cot ^2\beta}{8 v_\chi^2 v^2+v^2(1-\cos4\beta)}+m^2_{h_1}\frac{ \mathtt{C}^2_2 \mathtt{S}^2_3 \csc ^2\beta}{2 v^2}\nn\\
    &+m^2_{h_2}\frac{ \csc ^2\beta \left(\mathtt{S}^2_1 \mathtt{S}^2_2 \mathtt{S}^2_3+2\mathtt{S}_1 \mathtt{C}_1 \mathtt{S}_2 \mathtt{C}_3\mathtt{S}_3+\mathtt{C}^2_1 \mathtt{C}^2_3\right)}{2
   v^2}\nn\\
   &+m^2_{h_3}\frac{ \csc ^2\beta \left(\mathtt{C}_1 \mathtt{S}_2 \left(\mathtt{C}_1 \mathtt{S}_2 \mathtt{S}^2_3-2\mathtt{S}_1 \mathtt{C}_3\mathtt{S}_3\right)+\mathtt{S}^2_1 \mathtt{C}^2_3\right)}{2
   v^2},
\end{align}
\begin{align}        
\lambda_{12}&=m^2_{a_1}\frac{8 
   v_\chi^2}{8 v_\chi^2 v^2+v^2(1-\cos4\beta)}+m^2_{h_1}\frac{ \mathtt{C}^2_2 \mathtt{S}_3 \mathtt{C}_3 \csc \beta \sec
   \beta}{v^2}\nn\\
   &+m^2_{h_2}\frac{ \csc \beta \sec \beta}{4 v^2}
   \left(4 \mathtt{C}_1\mathtt{S}_1 \mathtt{S}_2(\mathtt{C}_3^2-\mathtt{S}_3^2) \right.\nn\\
   &\left.-\mathtt{C}_3\mathtt{S}_3 \left(2 \mathtt{S}^2_1(\mathtt{C}_2^2-\mathtt{S}_2^2) +6 \mathtt{C}_1\mathtt{S}_1+1\right)\right)\nn\\
   &-m^2_{h_3}\frac{ \csc \beta \sec \beta}{4
   v^2}
   \left(4 \mathtt{C}_1\mathtt{S}_1
   \mathtt{S}_2(\mathtt{C}_3^2-\mathtt{S}_3^2)\right.\nn\\
   &\left.+\mathtt{C}_3\mathtt{S}_3 \left(2 \mathtt{C}^2_1(\mathtt{C}_2^2-\mathtt{S}_2^2) -3(\mathtt{C}_1^2-\mathtt{S}_1^2)+1\right)\right),
\end{align}
\begin{align}    
\lambda_{23}&=m^2_{a_1}\frac{8  \cos ^2\beta}{8 v_\chi^2 +v^2(1-\cos4\beta)}-m^2_{h_1}\frac{ \mathtt{S}_2 \mathtt{C}_2
   \mathtt{S}_3 \csc \beta}{v_\chi
   v}\nn\\
   &+m^2_{h_2}\frac{ \mathtt{C}_2 \csc \beta
   \left(2 \mathtt{S}^2_1 \mathtt{S}_2 \mathtt{S}_3+2\mathtt{C}_1\mathtt{S}_1 \mathtt{C}_3\right)}{2 v_\chi
   v}\nn\\
   &-m^2_{h_3}\frac{ \mathtt{C}_2 \csc \beta
   \left(2\mathtt{C}_1\mathtt{S}_1 \mathtt{C}_3-2 \mathtt{C}^2_1
   \mathtt{S}_2 \mathtt{S}_3\right)}{2 v_\chi
   v},
\end{align}
\begin{align}    
\lambda_{13}&=m^2_{a_1}\frac{8  \sin ^2\beta}{8 v_\chi^2 +v^2(1-\cos4\beta)}-m^2_{h_1}\frac{ \mathtt{S}_2 \mathtt{C}_2
   \mathtt{C}_3 \sec \beta}{v_\chi
   v}\nn\\
   &+m^2_{h_2}\frac{ \mathtt{S}_1 \mathtt{C}_2
   \sec \beta (\mathtt{S}_1 \mathtt{S}_2 \mathtt{C}_3-\mathtt{C}_1 \mathtt{S}_3)}{v_\chi
   v}\nn\\
   &+m^2_{h_3} \frac{\mathtt{C}_1 \mathtt{C}_2
   \sec \beta (\mathtt{C}_1 \mathtt{S}_2 \mathtt{C}_3+\mathtt{S}_1 \mathtt{S}_3)}{v_\chi
   v},
\end{align}
\begin{align}\label{parsolkap}    
\lambda_3&=m^2_{a_1} \frac{v^2 \sin ^2 2\beta}{v_\chi^2 \left(v^2 (\cos4\beta-1)-8
   v_\chi^2\right)}\nn\\
   &+m^2_{h_1}\frac{ \mathtt{S}^2_2}{2
   v_\chi^2}+m^2_{h_2}\frac{ \mathtt{S}^2_1 \mathtt{C}^2_2}{2 v_\chi^2}+m^2_{h_3}\frac{ \mathtt{C}^2_1 \mathtt{C}^2_2}{2
   v_\chi^2}.         
\end{align}

Furthermore, it is straightforward to express the massive states of the charged sector of the general $331$ model in terms of the parameters of Eq.~\ref{pot}.
The three rotations read
\be
h^-_i=\mathcal{R}^C_{ij} H^-_j,
\quad
h^A_i=\mathcal{R}^{A}_{ij} H^A_j,
\quad
h^B_i=\mathcal{R}^{B}_{ij} H^B_j,
\ee
where $\vec H^-=((\rho^+)^*,\,\eta^-)$, $\vec h^-=(h^-_{G_W},\,h^-_1)$, $\vec H^A=((\eta^{-A})^*,\,\chi^A)$, $\vec h^A=(h^A_{G_{V^A}},\,h^A_1)$, $\vec H^B=((\rho^{-B})^*,\,\chi^B)$ and $\vec h^B=(h^B_{G_{V^B}},\,h^B_1)$. The mass matrices for the singly charged, $A$-charged and $B$-charged states are
\begin{align}
\mathcal{M}^2_{h^\pm}&=\left(
\begin{array}{cc}
 \kappa \tan \beta v_\chi^2+\frac{1}{2} \zeta_{12} v^2 \sin ^2\beta & \kappa
   v_\chi^2+\frac{1}{2} \zeta_{12} v^2 \cos \beta \sin \beta \\
  \sim & \kappa \cot \beta
   v_\chi^2+\frac{1}{2} \zeta_{12} v^2 \cos ^2\beta \\
\end{array}
\right)\\
\mathcal{M}^2_{h^{\pm A}}&=\left(
\begin{array}{cc}
 \frac{1}{2} v_\chi^2 (\zeta_{23}+2 \kappa \cot \beta) & \frac{1}{2} v_\chi v (2 \kappa \cos
   \beta+\zeta_{23} \sin \beta) \\
 \sim & \frac{1}{2} v^2 \sin
   \beta (2 \kappa \cos \beta+\zeta_{23} \sin \beta) \\
\end{array}
\right)\\
\mathcal{M}^2_{h^{\pm B}}&=\left(
\begin{array}{cc}
 \frac{1}{2} v_\chi^2 (\zeta_{13}+2 \kappa \tan \beta) & \frac{1}{2} v_\chi v (2 \kappa \sin
   \beta+\zeta_{13} \cos \beta) \\
 \sim & \frac{1}{2} v^2 \cos
   \beta (2 \kappa \sin \beta+\zeta_{13} \cos \beta) \\
\end{array}
\right)
\end{align}
From the equations above, one obtains:
\begin{align}
\zeta_{12}&=\frac{2}{v^2} \left(m_{h_1^\pm}^2-\frac{8 m_{a_1}^2 v_\chi^2}{8 v_\chi^2+v^2-v^2\cos4\beta}\right),\\
\zeta_{23}&=\frac{2 m^2_{h_1^{\pm A}}}{v^2 \sin ^2\beta+v_\chi^2}-\frac{16 m_{a_1}^2 \cos ^2\beta}{8 v_\chi^2+v^2-v^2\cos4\beta},\\
\zeta_{13}&=\frac{2 m^2_{h_1^{\pm B}}}{v^2 \cos ^2\beta+v_\chi^2}-\frac{16 m_{a_1}^2 \sin ^2\beta}{8 v_\chi^2+v^2-v^2\cos4\beta}.
\end{align}

These expressions show how to trade the 10 parameters of Eq.~\ref{pot} with the 3 physical rotation angles and the 7 physical scalar masses.

\end{document}